\documentclass[
    reprint,
    amsmath,
    amssymb,
    aps,
    physrev,
    longbibliography,
]{revtex4-2}
\usepackage{graphicx}
\usepackage{bm}
\usepackage{mathtools}
\usepackage{booktabs}
\usepackage{microtype}
\usepackage{hyperref}

\hypersetup{
    colorlinks=false,
    pdfborder={0 0 0}
}

\allowdisplaybreaks[2]

\begin{document}

\preprint{APS/123-QED}

\title{Bogoliubov Dynamics and Zitterbewegung in Optical Dirac Systems}

\author{Lili Yang$^1$}
\email{yanglli7@mail.sysu.edu.cn}

\affiliation{
 $^1$School of Physics and Astronomy, Sun Yat-Sen University, Zhuhai 519082, China}

\date{\today}% It is always \today, today,

\begin{abstract}
We develop a bosonic Bogoliubov description of Zitterbewegung in
time-dependent optical-Dirac systems. Within the optical-Dirac
framework, the anomalous pair-creation and pair-annihilation terms of
a bosonic Bogoliubov Hamiltonian correspond to couplings between the
positive- and negative-frequency sectors, whose coherence underlies
the Zitterbewegung response. When the constitutive parameters vary in
time, the resulting moving-frame connection generates dynamical
interbranch coupling and positive--negative-frequency coherence.
Under periodic modulation, this coupling becomes resonant near
$\Omega\simeq2\omega_0$, leading to parametric amplification of the
coherence and consequently to an enhanced Zitterbewegung motion.
\end{abstract}

\maketitle

\section{Introduction}

The Dirac equation provides one of the most fundamental examples in
which the internal structure of a wave function has direct dynamical
consequences. In particular, the coexistence of positive- and
negative-frequency branches gives rise to Zitterbewegung (ZB), the
rapid oscillatory motion originally identified in relativistic
quantum mechanics \cite{Dirac1928,Schrodinger1930,Huang1952,Thaller1992}.
Although the characteristic frequency and length scales make the
effect difficult to observe for free electrons, Dirac-like dynamics
has emerged in a wide variety of effective systems, including
graphene and other condensed-matter platforms
\cite{Katsnelson2006,Rusin2007}, ultracold atoms
\cite{Vaishnav2008}, trapped ions \cite{Gerritsma2010}, and
photonic systems \cite{Unal1997,KOBE19997,Longhi2010,Dreisow2010,Sedov2018,Yang2026NonAbelianZitterbewegung}.
These studies have established a general physical picture of ZB:
an observable oscillatory motion can arise when an operator has
nonvanishing matrix elements between branches of opposite frequency
and the state contains the corresponding interbranch coherence.

Optical systems provide a particularly natural setting for exploring
this structure. Maxwell's equations can be reorganized in terms of
complex field variables closely related to the
Riemann--Silberstein vector and interpreted in terms of photon wave
functions
\cite{BialynickiBirula1996,Sipe1995,Hawton1999,Tamburini2008,
BialynickiBirula2013,Keller2007,Barnett2014}. In generic media, the
optical-Dirac formulation can further incorporate the constitutive
tensor into effective mass and momentum operators, providing a compact
description of helicity, anisotropy, and spin--orbit coupling
\cite{PhysRevA.106.043513,yang2025inducedberryconnectionphotonic,
yang2025spinorbitcouplinghelicalwaveguides}. More recently, the same
four-component structure has been used as the basis for a
second-quantized description in which the positive- and
negative-frequency mode amplitudes are promoted to bosonic operators
and organized into a Nambu-like representation
\cite{r1vc-w3dl}.

In the optical-Dirac formulation, the electromagnetic field naturally
contains positive- and negative-frequency branches. In an anisotropic
medium, these two sectors can be mixed by the constitutive tensor and
therefore coexist in the dressed medium eigenstates. Upon second
quantization, the positive- and negative-frequency sectors are
organized as the annihilation and creation components of a bosonic
Nambu spinor. Their coupling consequently takes the form of anomalous
terms analogous to $a^\dagger a^\dagger$ and $aa$, characteristic of
pair creation and annihilation in quadratic bosonic Bogoliubov Hamiltonians
\cite{Bogoliubov1947,Colpa1978,CavesSchumaker1985,
SchumakerCaves1985,Clerk2010,LeinSato2019}.
This correspondence naturally places the optical-Dirac dynamics
within a bosonic Bogoliubov--de Gennes (BdG) framework for studying
Zitterbewegung. In a static
anisotropic medium, however, the positive--negative-frequency mixing
can be absorbed into the dressed eigenbasis by a suitable canonical
transformation, which diagonalizes the corresponding static
Hamiltonian. The static dressing therefore does not by itself
generate a dynamical interbranch coherence.

A qualitatively different situation arises when the constitutive
parameters of the medium become time dependent. Time-dependent
electromagnetic media have long been known to enable frequency
conversion and energy exchange between the modulation and the optical
field \cite{Morgenthaler1958,Fante1971}. Recent developments in
temporally modulated materials and metamaterials have extended these
ideas to temporal refraction and reflection, parametric amplification,
nonreciprocal transport, and Floquet-like photonic dynamics
\cite{Lustig2018,Sharabi2021,Galiffi2022,Yin2022,Moussa2023,
Pendry2021}. In the present setting, temporal modulation can
dynamically restore the coupling between the positive- and
negative-frequency components of the Nambu spinor, thereby generating
the interbranch coherence that is absent in the diagonal static
basis. Since Zitterbewegung originates from the interference between
the two frequency branches, this dynamically generated coherence
provides a direct mechanism for driving and parametrically enhancing
the observable trembling motion.

In this work, we show that the positive- and negative-frequency
structure of the optical-Dirac equation provides a natural framework
for connecting Zitterbewegung with bosonic Bogoliubov dynamics in a
time-dependent anisotropic medium. In a static medium, the
quadrupolar components $Q_{\pm2}$ mix the two frequency sectors in a
fixed reference basis. This mixing is absorbed into dressed
eigenstates through the boost-like transformation
$S(\Lambda)\equiv W$, which diagonalizes the leading optical-Dirac
Hamiltonian. In the Nambu representation, $W$ is $J$-paraunitary and
has the canonical structure of a bosonic Bogoliubov transformation
\cite{Colpa1978,LeinSato2019,TesfayeEckardt2025}. The static dressed basis
therefore removes dynamical coupling between the two frequency
branches, so that Zitterbewegung requires a pre-existing interbranch
coherence.

When the constitutive parameters become time dependent, the situation
changes qualitatively. Although the instantaneous optical-Dirac
Hamiltonian remains diagonalizable by $W(t)$, the time dependence of
the basis generates the moving-frame connection
$-iW^{-1}\dot W$. Its off-diagonal components act as anomalous
Bogoliubov couplings and dynamically generate positive--negative-
frequency coherence. For periodic modulation of $Q_{\pm2}(t)$, this
coupling becomes resonant near $\Omega\simeq2\omega_0$, leading to
parametric amplification of the coherence and, consequently, of the
ZB velocity and centroid oscillation. Pair production and
Zitterbewegung are therefore physically distinct but are governed by
the same interbranch coherence: the temporal modulation generates and
amplifies it, while the off-diagonal velocity matrix elements convert
it into observable trembling motion. Throughout this work, we adopt natural units with $c=\hbar=1$.

\section{Optical-Dirac modes in a time-dependent anisotropic medium}
\label{sec:time-dependent-medium}

The positive- and negative-frequency photon wave functions
$\psi^{\pm}$, which may also be regarded as canonical field
variables, are defined in the helicity basis as \cite{KELLER20051,PhysRevA.106.043513}
\begin{equation}\label{wavefunction}
    \begin{aligned}
        \psi^+_{\perp}(\mathbf r)
        &=
        D_{\perp}(\mathbf r)
        +
        i\sigma_3 B_{\perp}(\mathbf r),
        \\
        \psi^-_{\perp}(\mathbf r)
        &=
        D_{\perp}(\mathbf r)
        -
        i\sigma_3 B_{\perp}(\mathbf r).
    \end{aligned}
\end{equation}
These two components describe the positive- and negative-frequency
branches of the optical-Dirac equation, respectively. Although they
are convenient independent components of the optical-Dirac
representation, they ultimately originate from the same underlying
electromagnetic field and are therefore related by the corresponding
complex-conjugation constraint.

We consider a spin-degenerate medium satisfying
$\mu^{-1}=\epsilon^{-1}$, for which helicity is conserved
\cite{FernandezCorbaton2013,yang2025spinorbitcouplinghelicalwaveguides}.
The constitutive tensor is written as
\begin{equation}
    \mu^{-1}
    =
    \epsilon^{-1}
    =
    \begin{pmatrix}
        Q_0 & Q_{+2} & Q_{+1}\\
        Q_{-2} & Q_0 & Q_{-1}\\
        Q_{-1} & Q_{+1} & q_0
    \end{pmatrix}.
    \label{eq:medium-tensor}
\end{equation}
The tensor decomposition contains two scalar components,
$Q_0=\frac{1}{2}(\epsilon^{-1}_{11}+\epsilon^{-1}_{22})$ and
$q_0=\epsilon^{-1}_{33}$, the vector components
$Q_{\pm1}=\frac{1}{\sqrt{2}}
(\epsilon^{-1}_{13}\mp i\epsilon^{-1}_{23})$, which form
$\mathbf q=\{Q_{+1},Q_{-1}\}$, and the quadrupolar components
$Q_{\pm2}=\frac{1}{2}
[(\epsilon^{-1}_{11}-\epsilon^{-1}_{22})
\mp i(\epsilon^{-1}_{12}+\epsilon^{-1}_{21})]$.
These sectors transform as scalar (spin-$0$), vector (spin-$1$),
and tensor (spin-$2$) components, respectively
\cite{PhysRevA.106.043513}. This decomposition is particularly useful
because different tensor sectors couple differently to the helicity
degrees of freedom and therefore lead to distinct mode-mixing
selection rules.

The corresponding optical-Dirac equation is\cite{PhysRevA.106.043513,yang2025inducedberryconnectionphotonic}
\begin{equation}
    i\frac{\partial\psi(\mathbf r,t)}{\partial t}
    =
    \left(
        \beta\hat m
        +
        \beta\boldsymbol{\alpha}\cdot\hat{\mathbf p}
    \right)
    \psi(\mathbf r,t),
    \label{eq:optical-Dirac-medium}
\end{equation}
where $\beta=\gamma^0$ and
$\boldsymbol{\alpha}=\gamma^0\boldsymbol{\gamma}$. The effective
mass operator is
\begin{equation}
    \hat m
    =
    Q_0\hat k_z
    +
    \frac{q_0}{2k_z}
    \hat{\mathbf k}_{\perp}^{\,2}
    -
    \mathbf q\cdot\hat{\mathbf k}_{\perp},
    \label{eq:m-medium}
\end{equation}
while the transverse momentum operators are
\begin{equation}
    \hat p_{\pm}
    =
    Q_{\pm2}\hat k_z
    -
    2Q_{\pm1}\hat k_{\pm}
    +
    \frac{q_0}{k_z}\hat k_{\pm}^{\,2},
    \qquad
    \hat p_z=0.
    \label{eq:p-medium}
\end{equation}
The scalar, vector, and tensor components of the constitutive tensor
therefore enter different parts of the optical-Dirac generator. In
particular, the quadrupolar components $Q_{\pm2}$ directly mix the
positive- and negative-frequency sectors and will play the central
role in the following discussion.

We first introduce the fixed four-component reference basis
\begin{equation}
    \mathbf e_+
    =
    \begin{pmatrix}
        1\\0\\0\\0
    \end{pmatrix},\,
    \mathbf e_-
    =
    \begin{pmatrix}
        0\\1\\0\\0
    \end{pmatrix},\,
    \widetilde{\mathbf e}_+
    =
    \begin{pmatrix}
        0\\0\\1\\0
    \end{pmatrix},\,
    \widetilde{\mathbf e}_-
    =
    \begin{pmatrix}
        0\\0\\0\\1
    \end{pmatrix}.
    \label{eq:vacuum-positive-basis}
\end{equation}
We denote the corresponding basis matrix by
$\mathcal E_0=(\mathbf e_+,\mathbf e_-,
\widetilde{\mathbf e}_+,\widetilde{\mathbf e}_-)$, which is simply
$I_4$ in the present representation. To leading order in the longitudinal momentum $k_z$, The photon field can be expanded in plane-wave modes
in a form analogous to the standard mode expansion
of the quantized Dirac field, $\psi(\mathbf r,t)\propto
a_s(\mathbf k)\mathbf e_s(\mathbf k)
+b_s^\dagger(\mathbf k)\widetilde{\mathbf e}_s(\mathbf k)$\cite{Peskin1995AnIT,Williams2022IntroductionTQ,r1vc-w3dl}.
Here $a_s^\dagger$ and $b_s^\dagger$ create excitations associated
with the positive- and negative-frequency branches, respectively,
and $s=\pm$ labels the two helicities.

The terminology ``antiphoton'' may be used for the negative-frequency
branch, but it should not be understood as introducing an independent
photon species. Owing to the complex-conjugation constraint
$\psi_+=C\psi_-$ inherited from the underlying electromagnetic
fields, the positive- and negative-frequency sectors are not
independent. A negative-frequency excitation may therefore be
reinterpreted as a positive-frequency photon propagating in the
opposite direction. This interpretation will become useful below when
the anomalous Bogoliubov terms are related to physical photon-pair
channels.

Because the electromagnetic field is bosonic, the mode operators
satisfy the canonical commutation relations
\cite{r1vc-w3dl,cohentannoudji1989PhotonsAA}
\begin{equation}
\begin{aligned}
    [a_s(\mathbf k),a_{s'}^\dagger(\mathbf k')]
    &=
    \delta_{ss'}\delta^{(3)}(\mathbf k-\mathbf k'),
    \\
    [b_s(\mathbf k),b_{s'}^\dagger(\mathbf k')]
    &=
    \delta_{ss'}\delta^{(3)}(\mathbf k-\mathbf k').
\end{aligned}
\label{commutation relation}
\end{equation}
It is therefore convenient to introduce the bosonic Nambu operator
\begin{equation}
    \Phi_{\mathbf k}
    =
    \begin{pmatrix}
        a_{\mathbf k,+}\\
        a_{\mathbf k,-}\\
        b_{\mathbf k,+}^{\dagger}\\
        b_{\mathbf k,-}^{\dagger}
    \end{pmatrix}.
    \label{eq:Phi-fixed}
\end{equation}
The commutation relations then take the compact form
\begin{equation}
\begin{aligned}
    \left[
        \Phi_i(\mathbf k),
        \Phi_j^\dagger(\mathbf k')
    \right]
    &=
    J_{ij}
    \delta^{(3)}
    \left(
        \mathbf k-\mathbf k'
    \right),
    \\
    J
    &=
    \gamma^0
    =
    \operatorname{diag}(1,1,-1,-1).
\end{aligned}
\label{eq:Phi-commutator}
\end{equation}
The indefinite metric $J$ is the standard
canonical metric of bosonic BdG/Nambu systems and underlies their
paraunitary and Krein-space structure
\cite{Colpa1978,LeinSato2019,Flynn2020NJP,Massarelli2022}.
In the present optical-Dirac representation, this metric coincides
with $\gamma^0$ \cite{r1vc-w3dl}. The corresponding indefinite inner
product is therefore defined as
$\langle \psi_1,\psi_2\rangle_{\gamma^0}
=\langle \psi_1,\gamma^0\psi_2\rangle$.
This correspondence provides a natural bridge between the
single-particle optical-Dirac formulation and the bosonic BdG
description developed below.

\subsection{Instantaneous medium eigenbasis}

We now allow the quadrupolar components $Q_{\pm2}$ to vary in time.
For the construction of the instantaneous basis, we first focus on
the leading longitudinal sector governed by $Q_0$ and $Q_{\pm2}$.
The terms involving $Q_{\pm1}$ and finite transverse momentum
$k_\perp$ will be restored later as transverse corrections.

At each instant, the eigenstates of the anisotropic
medium can be related to the fixed free-photon
reference states through a time-dependent transformation
\cite{r1vc-w3dl}. We introduce this transformation
in the form of an optical boost-like operator,
\begin{equation}
    S[\Lambda(t)]
    \equiv
    W(t)
    =
    \exp\left[
        -\eta(t)\mathcal K(\phi)
    \right],
    \label{eq:boost-operator}
\end{equation}
where
\begin{equation}
    \mathcal K(\phi)
    =
    \begin{pmatrix}
        0&R(\phi)\\
        R(\phi)&0
    \end{pmatrix},\quad R(\phi)
    =
    \begin{pmatrix}
        0&e^{i\phi}\\
        e^{-i\phi}&0
    \end{pmatrix}.
    \label{eq:R-boost}
\end{equation}
Acting on the fixed reference states, this transformation
generates the instantaneous medium eigenstates
\cite{r1vc-w3dl},
\begin{equation}
\begin{aligned}
    \mathbf e_{\pm}(\mathbf k,t)
    =
    W(t)\,
    \mathbf e_{\pm}(k_z),\quad
    \widetilde{\mathbf e}_{\pm}(\mathbf k,t)
    =
    W(t)\,
    \widetilde{\mathbf e}_{\pm}(k_z).
\end{aligned}
\label{eq:boostrelation}
\end{equation}
For compactness, we suppress the explicit time
dependence of the medium parameters hereafter.
The corresponding eigenvectors take the form
\begin{equation}
\begin{aligned}
    \mathbf e_+(\mathbf{k})
    &=
    \mathcal N
    \begin{pmatrix}
        \lambda+Q_0\\
        0\\
        0\\
        -Q_{-2}
    \end{pmatrix},
    &
    \mathbf e_-(\mathbf{k})
    &=
    \mathcal N
    \begin{pmatrix}
        0\\
        \lambda+Q_0\\
        -Q_{+2}\\
        0
    \end{pmatrix},
    \\[3mm]
    \widetilde{\mathbf e}_-(\mathbf{k})
    &=
    \mathcal N
    \begin{pmatrix}
        -Q_{+2}\\
        0\\
        0\\
        \lambda+Q_0
    \end{pmatrix},
    &
    \widetilde{\mathbf e}_+(\mathbf{k})
    &=
    \mathcal N
    \begin{pmatrix}
        0\\
        -Q_{-2}\\
        \lambda+Q_0\\
        0
    \end{pmatrix}.
\end{aligned}
\label{eq:medium-positive-eigenvectors}
\end{equation}
Here
\begin{equation}
\begin{aligned}
    \mathcal N
    =
    \frac{k_z}
    {\sqrt{[\lambda+Q_0]k_z}},\quad
    \lambda^2=
    Q_0^2-|Q_2|^2.
\end{aligned}
\label{eq:dispersion}
\end{equation}
The quadrupolar anisotropy is characterized by
$Q_{\pm2}=Q_2e^{\pm i\phi}$, where $\phi$ specifies
its orientation. The strength of the mixing between
the positive- and negative-frequency sectors is
determined by the hyperbolic parameter $\eta(t)$,
which satisfies \cite{r1vc-w3dl}
\begin{equation}
    \tanh\eta(t)
    =
    \frac{Q_2(t)}
    {Q_0(t)+\lambda(t)},
    \label{eq:eta}
\end{equation}
Thus, the boost-like transformation is determined
directly by the constitutive parameters of the medium.

Since $R^\dagger=R$ and $R^2=I_2$, the generator
satisfies $\mathcal K^\dagger=\mathcal K$ and
$\mathcal K^2=I_4$. The transformation matrix
can therefore be evaluated explicitly as
\begin{equation}
    W(t)
    =
    \begin{pmatrix}
        \cosh\eta\,I_2
        &
        -\sinh\eta\,R(\phi)
        \\
        -\sinh\eta\,R(\phi)
        &
        \cosh\eta\,I_2
    \end{pmatrix}.
    \label{eq:boost-matrix}
\end{equation}
The hyperbolic structure of $W(t)$ is formally
analogous to that of a Lorentz boost acting on
a Dirac spinor, while here it describes the
medium-induced dressing of the free-photon
reference states.

Moreover, the generator $\mathcal K$ anticommutes
with the Nambu metric $J=\gamma^0$, ensuring that
the transformation preserves the corresponding
indefinite inner product,
\begin{equation}
    W^\dagger JW
    =
    J,
    \qquad
    W^{-1}
    =
    JW^\dagger J.
    \label{eq:paraW}
\end{equation}
Thus, $W$ is $J$-paraunitary rather than
ordinarily unitary, exhibiting the canonical
structure of a bosonic Bogoliubov transformation
in a Krein space
\cite{Colpa1978,LeinSato2019,Massarelli2022}.

Let $H_0$ denote the leading $Q_0$--$Q_{\pm2}$ part of the
optical-Dirac Hamiltonian. In the instantaneous basis defined in
Eq.~\eqref{eq:medium-positive-eigenvectors}, the $J$-paraunitary
transformation $W(t)$ diagonalizes $H_0(t)$ in the corresponding
Krein space, in close analogy with the paraunitary diagonalization
of bosonic BdG dynamical matrices
\cite{Colpa1978,LeinSato2019,Flynn2020NJP,Massarelli2022},
\begin{equation}
    W^{-1}(t)H_0(t)W(t)
    =
    \omega_0(t)J,
    \qquad
    \omega_0(t)
    =
    \lambda(t)k_z.
    \label{eq:HD-diagonal}
\end{equation}
The instantaneous spectrum therefore consists of two degenerate
positive-frequency states with eigenfrequency $+\omega_0(t)$ and two
degenerate negative-frequency states with eigenfrequency
$-\omega_0(t)$. Their separation is $2\omega_0(t)$, which later sets
the intrinsic oscillation frequency of the interbranch coherence.

We introduce the instantaneous basis matrix through
$\mathcal E(t)=
(\mathbf e_+(\mathbf k,t),\mathbf e_-(\mathbf k,t),
\widetilde{\mathbf e}_+(\mathbf k,t),
\widetilde{\mathbf e}_-(\mathbf k,t))$, so that
$\mathcal E(t)=\mathcal E_0W(t)$. The same field operator can then be
expanded either in the fixed reference basis or in the instantaneous
medium eigenbasis,
\begin{equation}
    \widehat{\Psi}_{\mathbf k}
    =
    \mathcal E_0\Phi_{\mathbf k}
    =
    \mathcal E(t)\Gamma_{\mathbf k}.
    \label{eq:two-expansions}
\end{equation}
The instantaneous Nambu operators are collected into the spinor
\begin{equation}
    \Gamma_{\mathbf k}
    =
    \begin{pmatrix}
        \alpha_{\mathbf k,+}\\
        \alpha_{\mathbf k,-}\\
        \beta_{\mathbf k,+}^{\dagger}\\
        \beta_{\mathbf k,-}^{\dagger}
    \end{pmatrix}.
    \label{eq:Gamma-instantaneous}
\end{equation}
Here, $\alpha_{\mathbf k,s}$ and
$\beta_{\mathbf k,s}^{\dagger}$ denote the quasiparticle annihilation
and creation operators associated with the instantaneous
positive- and negative-frequency modes of the medium, respectively.
Since the physical field itself is independent of the choice of
basis, the corresponding operator coordinates satisfy
\begin{equation}
    \Phi_{\mathbf k}
    =
    W(t)\Gamma_{\mathbf k},
    \qquad
    \Gamma_{\mathbf k}
    =
    W^{-1}(t)\Phi_{\mathbf k}.
    \label{eq:Phi-W-Gamma}
\end{equation}
At the level of Nambu space, the boost-like transformation $W$ is
$J$-paraunitary and therefore has the canonical structure of a
bosonic Bogoliubov transformation
\cite{Colpa1978,Ruijsenaars1978,LeinSato2019}.
Using Eqs.~\eqref{eq:Phi-commutator},
\eqref{eq:Phi-W-Gamma}, and \eqref{eq:paraW}, one obtains
\begin{equation}
    [\Gamma_i(\mathbf k),\Gamma_j^\dagger(\mathbf k^{\prime})]
    =
    J_{ij}
    \delta^{(3)}(\mathbf k-\mathbf k^{\prime}).
    \label{eq:GammaComm}
\end{equation}
Thus, the instantaneous quasiparticle operators obey the same
canonical commutation structure as the fixed-basis operators.

\section{Time-dependent instantaneous basis and Bogoliubov dynamics}
\label{sec:moving-frame}
Upon second quantization, the positive- and negative-frequency mode
amplitudes of the optical-Dirac field are promoted to bosonic
operators and assembled into the Nambu spinor $\Phi_{\mathbf k}$\cite{r1vc-w3dl,Colpa1978,LeinSato2019}.
The single-particle optical-Dirac generator then determines the
corresponding bosonic dynamical matrix in the fixed reference basis
defined in Eq.~\eqref{eq:vacuum-positive-basis}, so that
\begin{equation}
    i\partial_t\Phi_{\mathbf k}
    =
    D_{\rm ref}(t)\Phi_{\mathbf k}.
    \label{eq:Dref}
\end{equation}
Using the relation in Eq.~\eqref{eq:Phi-W-Gamma}, the dynamics in the
moving instantaneous quasiparticle basis is obtained as\cite{TesfayeEckardt2025}
\begin{equation}
    i\partial_t\Gamma_{\mathbf k}
    =
    \left[
        W^{-1}D_{\rm ref}W
        -
        iW^{-1}\dot W
    \right]
    \Gamma_{\mathbf k}.
    \label{eq:moving}
\end{equation}
The first term gives the instantaneous quasiparticle spectrum, $W^{-1}D_{\rm ref}W=D_0=\omega_0J$ , whereas the second term arises from the time dependence of the
instantaneous basis and acts as a moving-frame geometric connection. The full dynamical matrix in the
time-dependent instantaneous basis can therefore be written as
\begin{equation}
    D_{\rm inst}(t)
    =
    D_0(t)+\mathcal A_t(t),
    \qquad
    \mathcal A_t(t)
    =
    -iW^{-1}(t)\dot W(t).
    \label{eq:Dinst}
\end{equation}
The term $\mathcal A_t$ is the moving-frame geometric connection
generated by the temporal variation of the instantaneous eigenbasis.
Its diagonal components contribute to geometric phase accumulation,
whereas its off-diagonal components describe nonadiabatic transitions
between different instantaneous branches\cite{Berry1984,Rigolin2008}.

Differentiating the paraunitary condition in
Eq.~\eqref{eq:paraW} gives
\begin{equation}
    \left(
        W^{-1}\dot W
    \right)^\dagger J
    +
    J
    \left(
        W^{-1}\dot W
    \right)
    =
    0,
    \qquad
    \mathcal A_t^\dagger J
    =
    J\mathcal A_t.
    \label{eq:AJPseudo}
\end{equation}
Thus $\mathcal A_t$ is $J$-pseudo-Hermitian, as required for a
bosonic BdG dynamical matrix. The Hermitian quadratic Hamiltonian matrix corresponding to
$D_{\rm inst}$ is obtained through the bosonic Nambu metric $J$,
\begin{equation}
    \mathcal H_{\rm B}(t)
    =
    JD_{\rm inst}(t)
    =
    \omega_0(t)I_4
    +
    J\mathcal A_t(t).
    \label{eq:HBgeneral}
\end{equation}
Using Eq.~\eqref{eq:AJPseudo}, one readily verifies that
$\mathcal H_{\rm B}^\dagger=\mathcal H_{\rm B}$. Thus
$D_{\rm inst}$ generates the Heisenberg dynamics, while $\mathcal H_{\rm B}$ defines the corresponding quadratic bosonic
Hamiltonian.

For a static medium, $W$ is time independent,
so that $\mathcal A_t=0$ and
$\mathcal H_{\rm B}=\mathcal H_0=\omega_0 I_4$.
A nonzero static $Q_{\pm2}$ may strongly modify the eigenstates and
eigenfrequencies through the transformation $W$, but this mixing is
already incorporated into the dressed quasiparticle basis and does
not continuously generate transitions between the positive- and
negative-frequency sectors.

With the Nambu convention adopted here, the corresponding quadratic
second-quantized Hamiltonian operator can be written as
\begin{equation}
    \widehat H_0
    =
    \Gamma_{\mathbf k}^\dagger
    \mathcal H_0
    \Gamma_{\mathbf k}
    -
    2\omega_0,
\end{equation}
Using the bosonic commutation relation
$\beta_s\beta_s^\dagger=\beta_s^\dagger\beta_s+1$,
the Hamiltonian can be written in normal-ordered form as
\begin{equation}
    \widehat H_0
    =
    \omega_0
    \left(
        \alpha_+^\dagger\alpha_+
        +
        \alpha_-^\dagger\alpha_-
        +
        \beta_+^\dagger\beta_+
        +
        \beta_-^\dagger\beta_-
    \right).
    \label{eq:H0-operator}
\end{equation}
Equation~\eqref{eq:H0-operator} has the standard harmonic-oscillator
structure of quantized electromagnetic modes: each dressed
quasiparticle excitation carries the energy $\omega_0$.
Thus, in a static medium, the effect of $Q_{\pm2}$ is to dress the
instantaneous eigenmodes rather than to generate quasiparticle
conversion. Such dynamical interbranch coupling arises only when the
medium parameters, and consequently $W(t)$, vary in time.

\subsection{Time-dependent quadrupolar anisotropy}

We now consider a time-dependent quadrupolar anisotropy. For
simplicity, we first keep the anisotropy axis fixed,
$\dot\phi=0$, and allow only the magnitude $Q_2(t)$ to vary. The time
dependence of $W(t)$ is then carried entirely by $\eta(t)$.
Direct differentiation gives
\begin{equation}
    \mathcal A_t
    =
    i\dot\eta
    \begin{pmatrix}
        0&R(\phi)\\
        R(\phi)&0
    \end{pmatrix}.
    \label{eq:Aconstantphi}
\end{equation}
For this amplitude modulation, the moving-frame connection is purely
off-diagonal between the positive- and negative-frequency sectors.
It therefore acts as an interbranch Bogoliubov coupling rather than
as a diagonal geometric-phase contribution.

Substituting Eq.~\eqref{eq:Aconstantphi} into
Eq.~\eqref{eq:HBgeneral} gives the four-component bosonic BdG
Hamiltonian matrix
\begin{equation}
    \mathcal H_{\rm B}
    =
    \begin{pmatrix}
        \omega_0&0&0&g_+\\
        0&\omega_0&g_-&0\\
        0&g_-^*&\omega_0&0\\
        g_+^*&0&0&\omega_0
    \end{pmatrix},
    \label{eq:HB4}
\end{equation}
where
\begin{equation}
    g_+(t)
    =
    i\dot\eta(t)e^{i\phi},
    \qquad
    g_-(t)
    =
    i\dot\eta(t)e^{-i\phi}.
    \label{eq:geta}
\end{equation}
The anomalous coupling amplitudes are therefore not independent
phenomenological parameters: they are generated by the temporal
variation of the instantaneous medium eigenbasis.

Using Eq.~\eqref{eq:eta}, the time derivative of the boost parameter
is
\begin{equation}
    \dot\eta(t)
    =
    \frac{
        Q_0(t)\dot Q_2(t)-Q_2(t)\dot Q_0(t)
    }{
        2\lambda^2(t)
    }.
    \label{eq:eta-dot}
\end{equation}
In particular, when $Q_0$ is time independent and the phase $\phi$
is fixed, Eq.~\eqref{eq:geta} reduces to
\begin{equation}
    g_{\pm}(t)
    =
    \frac{iQ_0}{2\lambda^2(t)}
    \dot Q_{\pm2}(t).
    \label{eq:g-Q2}
\end{equation}
This relation makes an important distinction explicit: a static
quadrupolar anisotropy determines the dressed instantaneous
eigenstates, whereas its temporal variation generates the anomalous
Bogoliubov coupling. In particular, $Q_{\pm2}\neq0$ by itself does
not imply pair production, while $\dot Q_{\pm2}\neq0$ produces a
finite $g_\pm(t)$.

The corresponding second-quantized Hamiltonian is
\begin{equation}
    \widehat H_{\mathbf k}^{\rm B}
    =
    \sum_{s=\pm}
    \left[
        \omega_0
        \left(
            \alpha_s^\dagger\alpha_s
            +
            \beta_s^\dagger\beta_s
        \right)
        +
        g_s\alpha_s^\dagger\beta_{-s}^\dagger
        +
        g_s^*\alpha_s\beta_{-s}
    \right].
    \label{eq:HBoperator}
\end{equation}
The terms proportional to $\omega_0$ describe the instantaneous
quasiparticle energies, whereas those proportional to $g_s$ have the
standard anomalous structure of a bosonic Bogoliubov Hamiltonian
\cite{Bogoliubov1947,Colpa1978}. In the present system, these anomalous terms are generated by the
off-diagonal moving-frame connection associated with the
time-dependent medium and couple the positive- and negative-frequency
sectors.

The Hermitian quadratic matrix
$\mathcal H_{\rm B}$ associated with
Eq.~\eqref{eq:HBoperator} determines the energy functional, whereas
the corresponding Heisenberg evolution is generated by the bosonic
BdG dynamical matrix
$D_{\rm B}=J\mathcal H_{\rm B}$.
For the present $Q_{\pm2}$ tensor sector, the four-component dynamics
separates naturally into two anomalous channels,
$(\alpha_+,\beta_-^\dagger)^T$ and
$(\alpha_-,\beta_+^\dagger)^T$.
Each channel is described by a standard $2\times2$ bosonic BdG
dynamical matrix, and the explicit reduction is given in
Appendix~\ref{app:two-BdG-channels}.

\subsection{Physical interpretation and polarization selection rule}

We next discuss the physical interpretation of the anomalous terms in
Eq.~\eqref{eq:HBoperator}. Even when the constitutive parameters vary
in time, the reality condition of the underlying electromagnetic
fields continues to relate the positive- and negative-frequency
components. The $\beta$ sector therefore does not represent an
independent photon species. Within the branch convention adopted
here, a negative-frequency mode labeled by $(\mathbf k,s)$ can be
reinterpreted as a positive-frequency photon propagating with the
opposite momentum. For the purpose of identifying the corresponding
physical photon modes, we therefore use the correspondence
\begin{equation}
    \beta_{\mathbf k,s}
    \leftrightarrow
    \alpha_{-\mathbf k,s}.
    \label{eq:beta-alpha-identification}
\end{equation}
Under this correspondence, the anomalous creation terms in
Eq.~\eqref{eq:HBoperator} are interpreted as
\begin{equation}
    \alpha_{\mathbf k,\pm}^{\dagger}
    \beta_{\mathbf k,\mp}^{\dagger}
    \longrightarrow
    \alpha_{\mathbf k,\pm}^{\dagger}
    \alpha_{-\mathbf k,\mp}^{\dagger},
    \label{eq:physical-pair-channels}
\end{equation}
The negative-frequency branch thus plays the role of the hole-like
Nambu component, and the quasiparticle dynamics acquires the standard
bosonic Bogoliubov--de Gennes structure
\cite{Colpa1978,LeinSato2019,Flynn2020EPL}.
Together with their Hermitian-conjugate counterparts, the anomalous
terms describe the creation and annihilation of correlated photon
pairs, analogous to two-mode parametric processes in quantum optics
\cite{CavesSchumaker1985,SchumakerCaves1985,WallsMilburn2008,
Ozawa2019TopologicalPhotonics}.

The relation between physical helicity and spin projection on a fixed
laboratory axis requires some care. For a photon propagating along
$+\hat{\mathbf z}$, the fixed-axis spin projection is $S_z=s$,
whereas for propagation along $-\hat{\mathbf z}$ it is $S_z=-s$,
because helicity is defined relative to the direction of propagation.
Consequently, although the two photons in each pair carry opposite
helicity labels, their spin projections on the fixed $z$ axis have
the same sign. In the leading longitudinal sector, the two pairing
channels therefore satisfy
\begin{equation}
    (\mathbf k,\pm)+(-\mathbf k,\mp)
    \quad\Rightarrow\quad
    \Delta S_z=\pm2.
    \label{eq:spin-selection}
\end{equation}
When helicity is conserved in a spin-degenerate medium, the
corresponding helicity sectors evolve independently
\cite{FernandezCorbaton2013,yang2025spinorbitcouplinghelicalwaveguides}.
This selection rule reflects the rank-$2$ angular-momentum
character of the quadrupolar components $Q_{\pm2}$, whose
angular-momentum indices with respect to the fixed $z$ axis are
$\pm2$. Accordingly, the interaction terms involving $Q_{\pm2}$
couple to the pair channels
$\alpha_s^\dagger\beta_{-s}^\dagger$ in
Eq.~\eqref{eq:HBoperator}, describing angular-momentum exchange
between the time-dependent medium and the optical field.
In the absence of external torque, total angular momentum is
conserved when the medium and the source of modulation are
included, whereas the optical subsystem may gain or lose angular
momentum through the modulation.

\section{Zitterbewegung and Its Parametric Enhancement}
\label{sec:ZB}

The Bogoliubov dynamics derived above provides a natural setting for
examining Zitterbewegung in the optical-Dirac system. We first
consider the static case, where ZB originates from the coherence
between the positive- and negative-frequency branches. We then
restore the temporal modulation of $Q_{\pm2}(t)$ and show that the
resulting Bogoliubov coupling dynamically generates and amplifies
this interbranch coherence. Near the parametric-resonance condition
$\Omega\simeq2\omega_0$, the interbranch coherence
is strongly enhanced, resulting in an amplified
ZB-related velocity response and a corresponding
effective oscillatory displacement.

\subsection{ZB velocity and effective displacement}

As derived in Appendix~B, the optical-Dirac velocity
operator follows from
$\hat v_\mu=i[\hat H_D,\hat r_\mu]$.
In the instantaneous medium eigenbasis, the
transformed velocity matrix $W^{-1}v_\mu W$
contains off-diagonal elements $C_{s,\mu}$
connecting the positive- and negative-frequency
branches. In terms of the quasiparticle operators,
the corresponding second-quantized ZB velocity
contribution takes the form
\begin{equation}
    \widehat v_\mu^{\rm ZB}
    =
    \sum_{s=\pm}
    \left[
        C_{s,\mu}
        \alpha_s^\dagger\beta_{-s}^\dagger
        +
        C_{s,\mu}^*
        \alpha_s\beta_{-s}
    \right],
    \label{eq:vZB-operator}
\end{equation}
where $\mu=+,-,z$, and the coefficients $C_{s,\mu}$ are given
explicitly in Appendix~B.
The ZB contribution is controlled by the coherence between the two
frequency sectors. We characterize this coherence by the anomalous
correlation
\begin{equation}
    M_s(t)
    \equiv
    \left\langle
        \alpha_s(t)\beta_{-s}(t)
    \right\rangle .
    \label{eq:Ms-definition}
\end{equation}
The corresponding expectation value of the ZB velocity is then
\begin{equation}
    \left\langle
        \widehat v_\mu^{\rm ZB}(t)
    \right\rangle
    =
    2\sum_{s=\pm}
    \operatorname{Re}
    \left[
        C_{s,\mu}^*M_s(t)
    \right].
    \label{eq:vZB-expectation}
\end{equation}
Thus, $C_{s,\mu}$ specifies how the velocity observable couples the
two frequency branches, whereas $M_s(t)$ measures the corresponding
interbranch coherence. A finite ZB signal therefore requires both a
nonvanishing off-diagonal velocity matrix element and a finite
positive--negative frequency coherence.

In the absence of temporal modulation, the quasiparticle operators
evolve according to the free Hamiltonian
Eq.~\eqref{eq:H0-operator}, giving
$M_s(t)=M_s(0)e^{-2i\omega_0t}$. The frequency $2\omega_0$ is the
separation between the positive- and negative-frequency branches.
A static medium does not generate this coherence: if $M_s(0)=0$, it
remains zero, whereas a pre-existing coherence undergoes free phase
evolution at the interbranch splitting. Substituting this result into
Eq.~\eqref{eq:vZB-expectation} gives
\begin{equation}
    \left\langle
        \widehat v_\mu^{\rm ZB}(t)
    \right\rangle
    =
    2\sum_{s=\pm}
    \operatorname{Re}
    \left[
        C_{s,\mu}^*M_s(0)e^{-2i\omega_0t}
    \right].
    \label{eq:vZB-free}
\end{equation}
The ZB velocity therefore oscillates at the
positive--negative-frequency splitting $2\omega_0$, with an amplitude
determined jointly by the interbranch velocity matrix element
$C_{s,\mu}$ and the initial coherence $M_s(0)$.

For a narrow wave packet centered at $\mathbf k_0$,
$C_{s,\mu}$ and $\omega_0$ may be evaluated at $\mathbf k_0$.
Within the effective narrow-wave-packet description,
we employ the Ehrenfest relation
$d\langle r_\mu\rangle/dt=\langle v_\mu\rangle$
to characterize the oscillatory motion associated
with the interbranch velocity.
Integrating Eq.~\eqref{eq:vZB-free} yields
the corresponding effective ZB displacement,
\begin{equation}
    \Delta r_\mu^{\rm ZB}(t)
    =
    \sum_{s=\pm}
    \frac{1}{\omega_0}
    \operatorname{Im}
    \left\{
        C_{s,\mu}^*M_s(0)
        \left[
            1-e^{-2i\omega_0t}
        \right]
    \right\}.
    \label{eq:rZB-free}
\end{equation}
The characteristic displacement amplitude of each channel is
therefore of order
$\left|C_{s,\mu}M_s(0)\right|/\omega_0$.

Equations~\eqref{eq:vZB-free} and \eqref{eq:rZB-free} make explicit
that the ZB signal arises from interference between the positive- and
negative-frequency components of the optical-Dirac field.
The coefficient $C_{s,\mu}$ determines how this coherence is probed by
the velocity operator, while $M_s(t)$ characterizes the coherence
itself. This distinction will be important below, where the temporal
modulation dynamically generates and amplifies $M_s(t)$.

\subsection{Parametrically enhanced Zitterbewegung}
We now restore the temporal modulation of the quadrupolar anisotropy.
Periodic modulation of constitutive parameters provides a natural
route to frequency conversion and parametric amplification in
time-varying photonic media. In particular, photonic time crystals
are characterized by a temporal periodicity of the electromagnetic
environment and exhibit Floquet-like mode coupling, frequency-band
formation, and parametric instabilities that have no direct analogue
in a static medium
\cite{Lustig2018,Sharabi2021,Galiffi2022,Yin2022}.
Related temporal interfaces and time-modulated metamaterials have also
been shown to support temporal reflection, frequency translation, and
gain through energy exchange between the modulation and the optical
field
\cite{PachecoPena2020,Pendry2021,Moussa2023}.

Here we consider a weak periodic modulation of the quadrupolar
component,
\begin{equation}
    Q_2(t)
    =
    \bar Q_2
    +
    \delta Q\cos\Omega t,
    \qquad
    |\delta Q|\ll Q_0,
    \label{eq:Qmod-ZB}
\end{equation}
with a fixed orientation $\phi$. The static part $\bar Q_2$
determines the instantaneous dressed eigenbasis, whereas the
oscillatory part $\delta Q\cos\Omega t$ drives transitions between
the positive- and negative-frequency sectors through the
time-dependent Bogoliubov coupling. Because these two sectors are
separated by approximately $2\omega_0$, the modulation becomes
parametrically resonant when $\Omega$ approaches this frequency
splitting. To leading order in the modulation amplitude, the
induced Bogoliubov coupling is
\begin{equation}
    g_{\pm}(t)
    \simeq
    -i e^{\pm i\phi}
    \frac{
        Q_0\delta Q\Omega
    }{
        2\bar\lambda^2
    }
    \sin\Omega t,
    \qquad
    \bar\lambda^2
    =
    Q_0^2-\bar Q_2^2.
    \label{eq:gmod-ZB}
\end{equation}
The temporal modulation therefore supplies an oscillatory anomalous
coupling between the positive- and negative-frequency sectors. Since
their separation is $2\omega_0$, this coupling becomes resonant when
its driving frequency approaches the same scale.
\begin{figure*}[t]
    \centering
    \begin{minipage}[t]{0.49\textwidth}
        \centering
        \includegraphics[width=\linewidth]{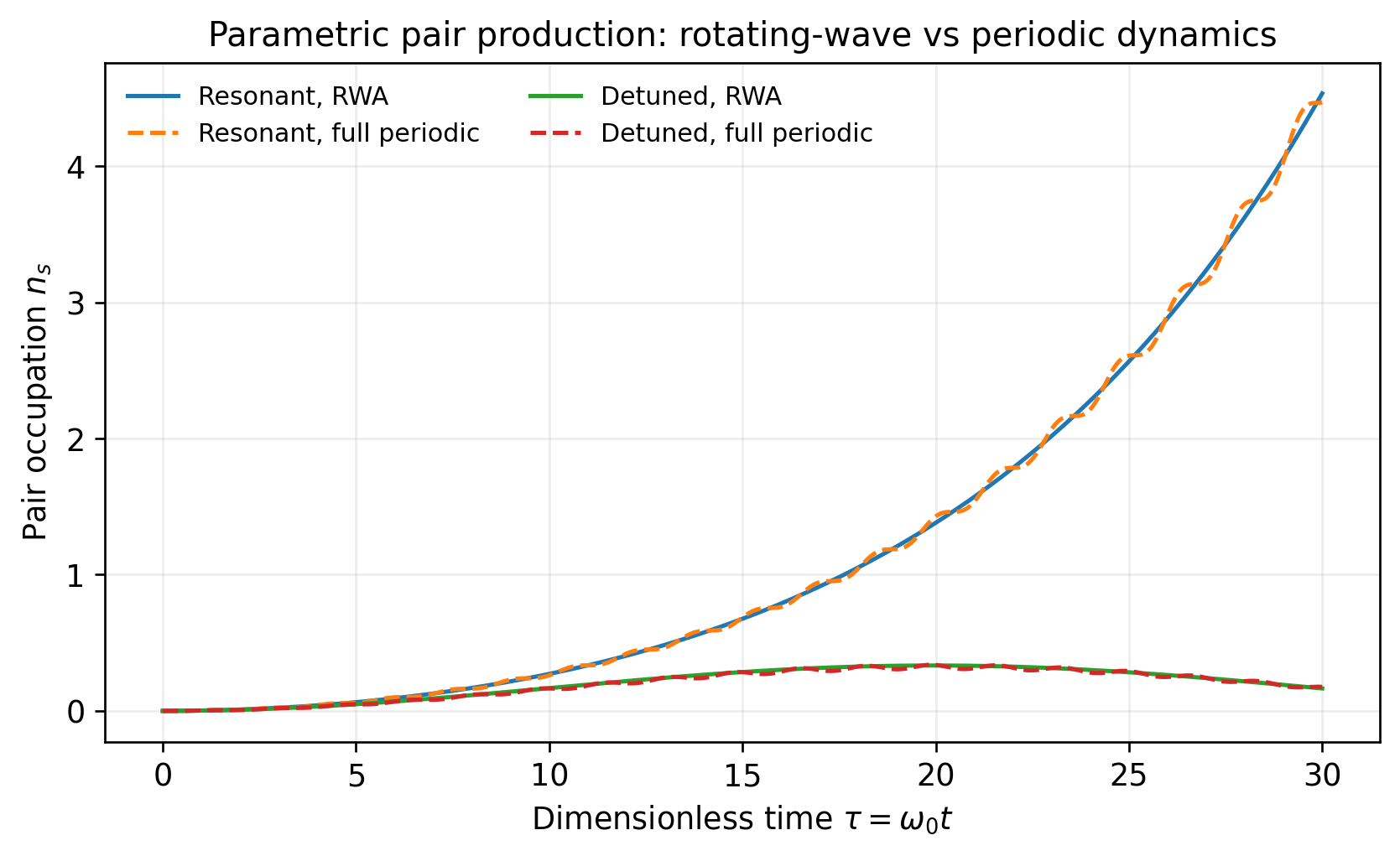}
        \par\small (a)
    \end{minipage}\hfill
    \begin{minipage}[t]{0.49\textwidth}
        \centering
        \includegraphics[width=\linewidth]{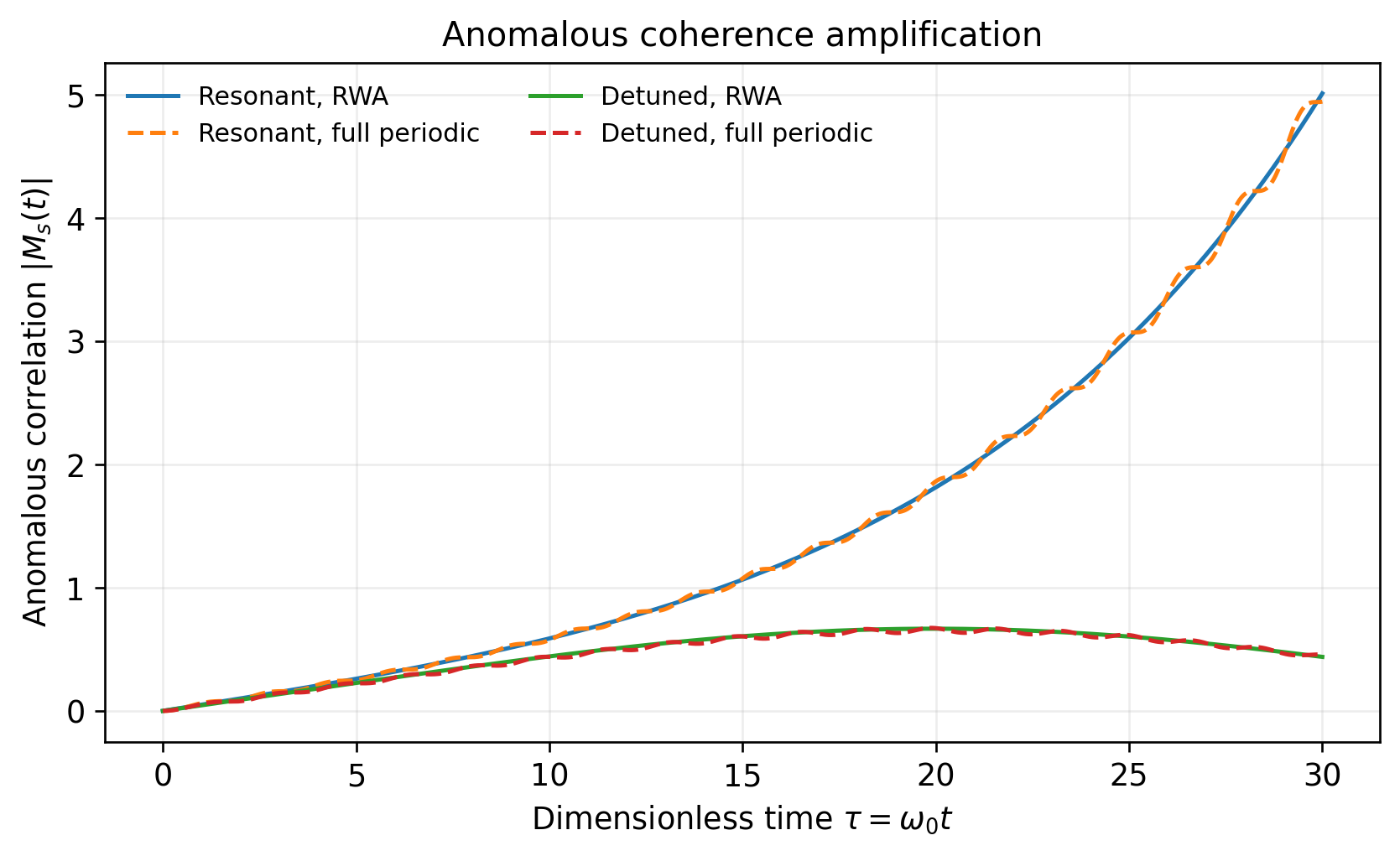}
        \par\small (b)
    \end{minipage}
    \caption{
        Numerical Bogoliubov dynamics starting from the quasiparticle
        vacuum: (a) pair occupation $n_s(t)$ and (b) anomalous
        coherence magnitude $|M_s(t)|$. Solid curves denote the
        rotating-wave approximation (RWA), while dashed curves are
        obtained by integrating the time-periodic coupling $g_s(t)$
        without applying the RWA. The resonant case has
        $\Omega=2\omega_0$ and $|G_s|/\omega_0=0.05$; the
        detuned case has $\Delta_s=2|G_s|$.
        Resonance produces parametric growth of both quantities,
        whereas the detuned solutions remain bounded.
        The small oscillatory deviations of the full-periodic results
        reflect counter-rotating contributions.
    }
    \label{fig:population-coherence}
\end{figure*}
In the weak-modulation regime, the velocity matrix elements
$C_{s,\mu}$ may be evaluated using the static reference basis
determined by $\bar Q_2$. The time-dependent coupling $g_s(t)$ then
acts as a source of positive--negative-frequency coherence, which is
most conveniently analyzed in the rotating frame near
$\Omega\simeq2\omega_0$. The intrinsic ZB frequency $2\omega_0$ therefore coincides with the
parametric-resonance condition of the driven Bogoliubov dynamics.
This coincidence does not mean that Bogoliubov pair production and
Zitterbewegung are the same phenomenon. The former generates
correlated excitations and anomalous coherence, whereas the latter is
the observable oscillatory response produced when this coherence is
sampled by the off-diagonal velocity matrix elements.

Near $\Omega\simeq2\omega_0$, it is useful to decompose the periodic
Bogoliubov coupling into its two Fourier components. Using
Eq.~\eqref{eq:gmod-ZB}, we write
\begin{equation}
    g_s(t)
    =
    G_s e^{-i\Omega t}
    -
    G_s e^{i\Omega t},
    \qquad
    G_s
    =
    \frac{Q_0\delta Q\Omega}
    {4\bar\lambda^2}
    e^{is\phi}.
    \label{eq:Gs-definition}
\end{equation}
Here $G_s$ is the amplitude of the Fourier component resonant with
pair creation. In the interaction picture, the pair-creation
operator $\alpha_s^\dagger\beta_{-s}^\dagger$ carries the free phase
$e^{2i\omega_0t}$. Hence the component proportional to
$e^{-i\Omega t}$ varies as
$e^{i(2\omega_0-\Omega)t}$ and becomes resonant near
$\Omega=2\omega_0$, whereas the counter-rotating component varies as
$e^{i(2\omega_0+\Omega)t}$ and is neglected. The remaining explicit time dependence is removed by introducing
operators in a frame rotating at $\Omega/2$,
$\alpha_s^{\rm lab}=e^{-i\Omega t/2}\alpha_s^{\rm rot}$ and
$\beta_{-s}^{\rm lab}=e^{-i\Omega t/2}\beta_{-s}^{\rm rot}$.
Accordingly, the anomalous coherence defined in
Eq.~\eqref{eq:Ms-definition} satisfies
\begin{equation}
    M_s(t)=e^{-i\Omega t} 
\langle
\alpha_s^{\rm rot}(t)
\beta_{-s}^{\rm rot}(t)
\rangle=e^{-i\Omega t}
    M_s^{\rm rot}(t) ,
    \label{eq:M-rot-relation}
\end{equation}
Thus, $M_s^{\rm rot}(t)$ describes the slowly varying Bogoliubov
dynamics, while the factor $e^{-i\Omega t}$ restores the physical
phase evolution of the coherence in the original frame.
For notational simplicity, the superscript ``rot'' on the
quasiparticle operators is omitted below, unless needed explicitly. 
The effective rotating-frame
Hamiltonian is
\begin{equation}
    \widehat H_{\mathrm{rot}}^{(s)}
    =
    \Delta_s
    \left(
        \alpha_s^{\dagger}\alpha_s
        +
        \beta_{-s}^{\dagger}\beta_{-s}
    \right)
    +
    G_s\alpha_s^{\dagger}\beta_{-s}^{\dagger}
    +
    G_s^*\alpha_s\beta_{-s},
    \label{eq:Hrot}
\end{equation}
where $\Delta_s=\omega_0-\Omega/2$ and
$G_s=|G_s|e^{i\theta_s}$. The corresponding dynamical eigenvalues are
\begin{equation}
    \varepsilon_{\mathrm{rot},\pm}
    =
    \pm
    \sqrt{
        \Delta_s^2-|G_s|^2
    }.
    \label{eq:rot-eigenvalues}
\end{equation}
When $|G_s|>|\Delta_s|$, the eigenvalues become imaginary and the
Bogoliubov amplitudes acquire exponentially growing components,
signaling the parametric-amplification regime.

At exact resonance, $\Delta_s=0$, the rotating-frame Hamiltonian
reduces to
\begin{equation}
    \widehat H_{\mathrm{rot}}^{(s)}
    =
    G_s\alpha_s^{\dagger}\beta_{-s}^{\dagger}
    +
    G_s^*\alpha_s\beta_{-s}.
    \label{eq:Hrot-resonance}
\end{equation}
The corresponding Heisenberg equations are
\begin{equation}
    i\dot{\alpha}_s
    =
    G_s\beta_{-s}^{\dagger},
    \qquad
    i\dot{\beta}_{-s}^{\dagger}
    =
    -G_s^*\alpha_s,
    \label{eq:Heisenberg-resonance}
\end{equation}
with solutions
\begin{align}
    \alpha_s(t)
    &=
    \cosh\left(|G_s|t\right)\alpha_s(0)
    -
    i e^{i\theta_s}
    \sinh\left(|G_s|t\right)
    \beta_{-s}^{\dagger}(0),
    \label{eq:alpha-solution}
    \\
    \beta_{-s}^{\dagger}(t)
    &=
    \cosh\left(|G_s|t\right)\beta_{-s}^{\dagger}(0)
    +
    i e^{-i\theta_s}
    \sinh\left(|G_s|t\right)
    \alpha_s(0).
    \label{eq:beta-dagger-solution}
\end{align}
For the driven dynamics, the rotating-frame anomalous coherence obeys
\begin{equation}
    i\dot M_s^{\rm rot}
    =
    G_s
    \left[
        1+n_{\alpha,s}+n_{\beta,-s}
    \right].
    \label{eq:Ms-rot-eom}
\end{equation}
\begin{figure*}[t]
    \centering
    \begin{minipage}[t]{0.49\textwidth}
        \centering
        \includegraphics[width=\linewidth]{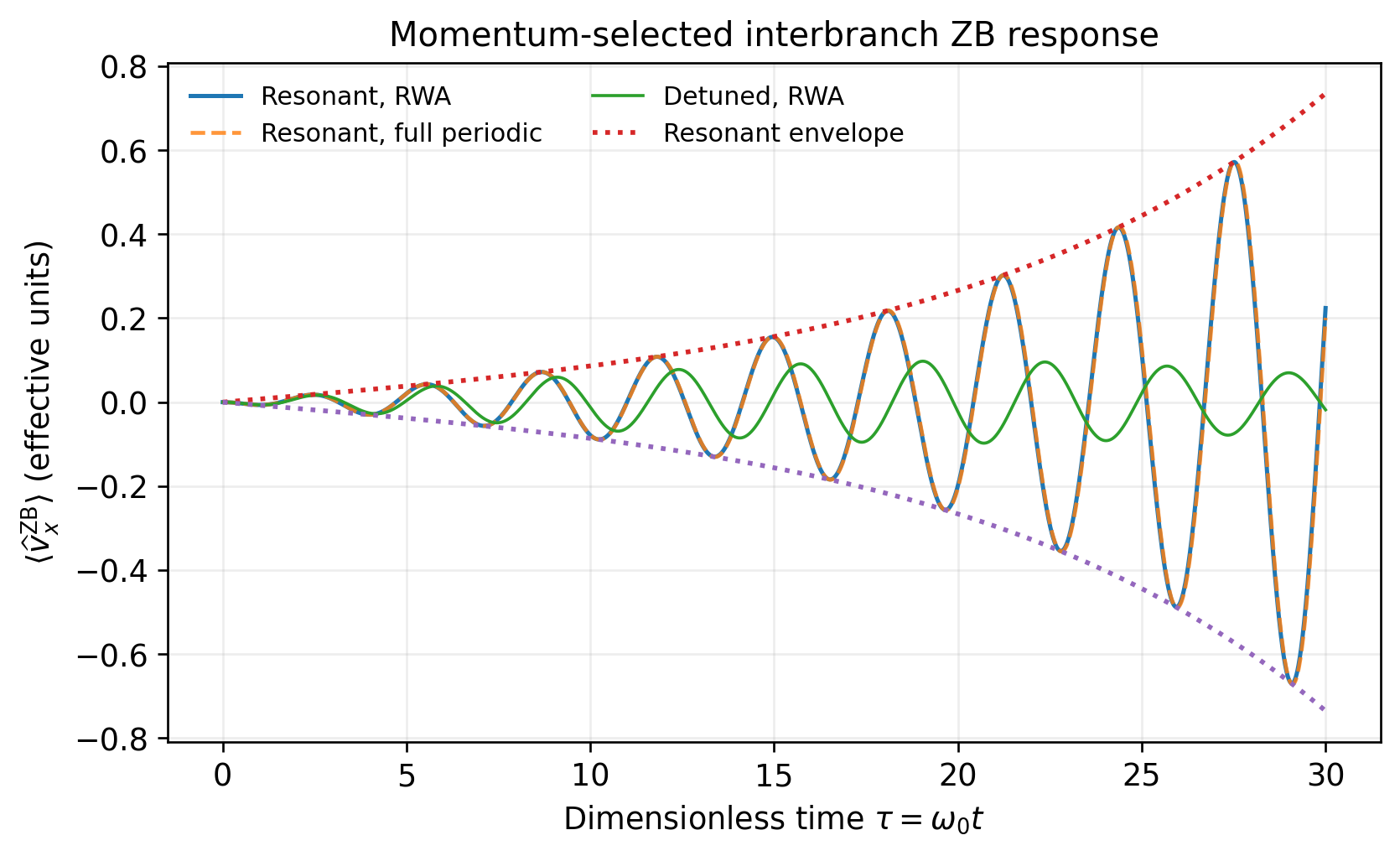}
        \par\small (a)
    \end{minipage}\hfill
    \begin{minipage}[t]{0.49\textwidth}
        \centering
        \includegraphics[width=\linewidth]{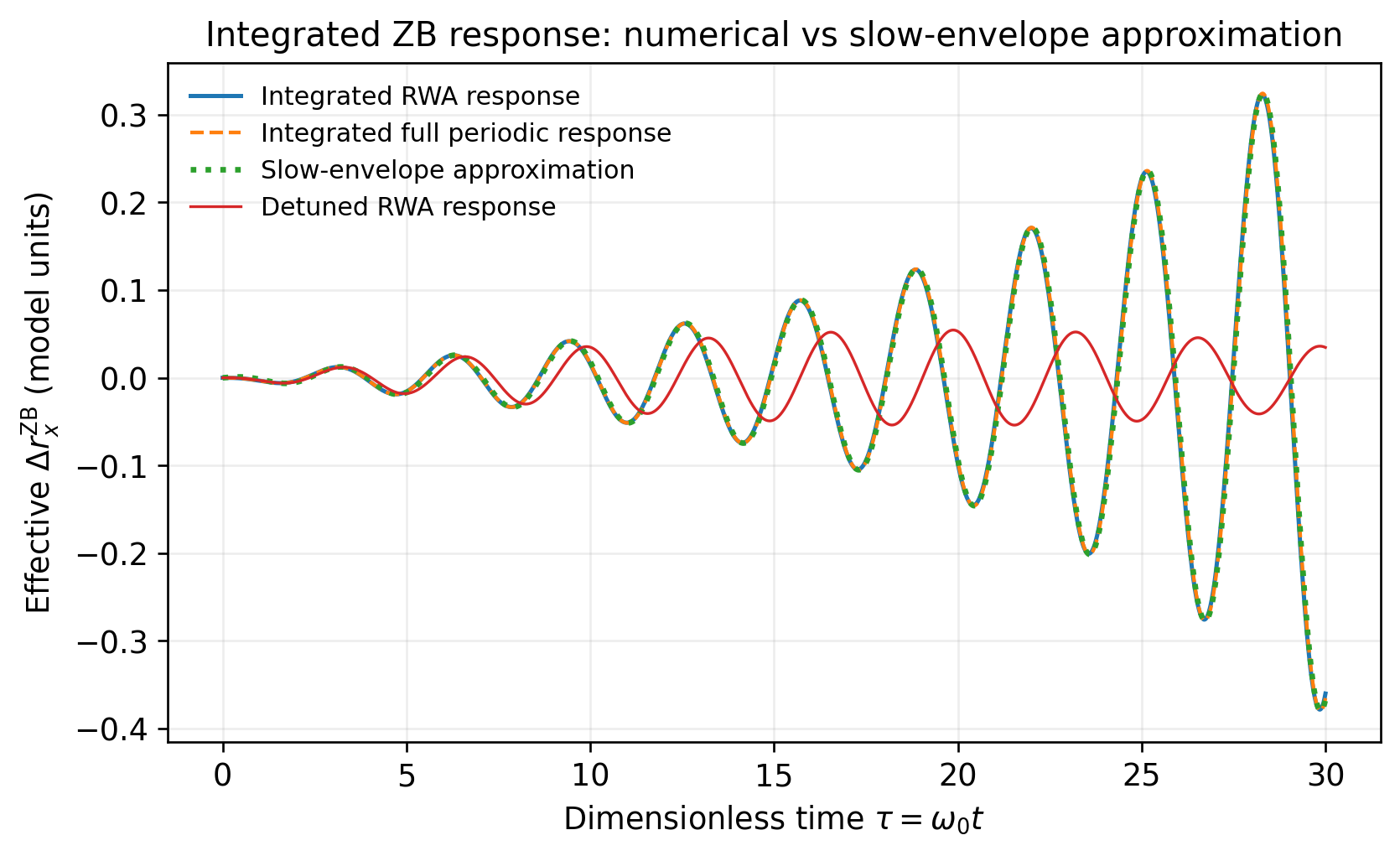}
        \par\small (b)
    \end{minipage}
    \caption{
        Numerical parametric enhancement of the momentum-selected,
        effective ZB response.
        (a) Interbranch contribution $\langle\widehat v_x^{\rm ZB}\rangle$
        evaluated from the anomalous correlation and the velocity
        matrix elements at $k_{x0}=0.25$, $k_{y0}=0$.
        The rotating-wave and full-periodic calculations closely
        agree at resonance; the detuned RWA response remains bounded.
        Dotted curves indicate the resonant amplitude envelope.
        (b) Corresponding effective ZB displacement obtained by
        integrating the oscillatory velocity response. The direct
        time integrations using the RWA and the full-periodic model
        are compared with the slow-envelope approximation of
        Eq.~\eqref{eq:rZB-resonant}, together with the detuned
        RWA result. The displacement is an effective, mode-resolved
        response and is not identified here with a separately
        normalized electromagnetic energy centroid.
    }
    \label{fig:ZB-enhancement}
\end{figure*}

Here,
$n_{\alpha,s}(t)=\langle\alpha_s^\dagger(t)\alpha_s(t)\rangle$ and
$n_{\beta,-s}(t)=\langle\beta_{-s}^\dagger(t)\beta_{-s}(t)\rangle$
are the corresponding quasiparticle occupation numbers. We take the initial state to be the instantaneous quasiparticle
vacuum, satisfying
$\alpha_s(0)|0\rangle=0$ and
$\beta_{-s}(0)|0\rangle=0$.
The exact Bogoliubov solutions then give equal populations in the two
modes,
\begin{equation}
    n_{\alpha,s}(t)
    =
    n_{\beta,-s}(t)
    \equiv
    n_s(t)
    =
    \sinh^2\left(|G_s|t\right).
    \label{eq:ns-rot}
\end{equation}
The equality of the two populations reflects the correlated
two-mode nature of the parametric excitation. The occupation number
characterizes the generated quasiparticle population, but does not by
itself determine the ZB signal.

Thus, for an initial quasiparticle vacuum and at exact resonance,
the rotating-frame anomalous coherence is
\begin{equation}
    M_s^{\rm rot}(t)
    =
    -\frac{i}{2}
    e^{i\theta_s}
    \sinh\left(2|G_s|t\right).
    \label{eq:Ms-rot}
\end{equation}
Consequently,
\begin{equation}
    \left|M_s^{\rm rot}(t)\right|^2
    =
    n_s(t)\left[n_s(t)+1\right].
    \label{eq:M-n-relation}
\end{equation}

To examine the regime of validity of the rotating-wave approximation
(RWA), we numerically integrate the two-mode Bogoliubov evolution both
in the RWA and with the complete periodic coupling $g_s(t)$ retained.
We use dimensionless model parameters $Q_0=1$,
$\bar Q_2=0.30$, $q_0=0.20$, $k_z=1$,
$\delta Q=0.091$, and $\phi=0$.
At resonance $\Omega=2\omega_0$, the parameters give
$|G_s|/\omega_0=0.05$; the detuned example is chosen so that
$\Delta_s=2|G_s|$. Time is plotted in units of
$\tau=\omega_0t$.

The numerical results are shown in
Fig.~\ref{fig:population-coherence}.
At resonance, both $n_s(t)$ and $|M_s^{\rm rot}(t)|$ grow
parametrically, whereas in the off-resonant regime
$|\Delta_s|>|G_s|$ they remain bounded.
Since $|M_s(t)|=|M_s^{\rm rot}(t)|$, the resonant growth of the
rotating-frame anomalous coherence determines the growing envelope
of the interbranch coherence, which provides the microscopic basis
of the enhanced ZB-related response. 

For the vacuum-generated
two-mode squeezed state, $n_s(t)$ and $M_s^{\rm rot}(t)$ are fixed by
the same Bogoliubov dynamics but represent different physical
quantities: the former measures the generated quasiparticle
population, whereas the latter characterizes the phase-sensitive
interbranch coherence. Substituting $M_s(t)$ into
Eq.~\eqref{eq:vZB-expectation} gives
\begin{equation}
    \left\langle
        v_{\mu}^{\rm ZB}(t)
    \right\rangle_{\rm res}
    =
    \sum_{s=\pm}
    |C_{s,\mu}|
    \sinh\left(2|G_s|t\right)
    \cos\left(
        2\omega_0t-\Phi_{s,\mu}
    \right),
    \label{eq:vZB-resonant}
\end{equation}
where
$\Phi_{s,\mu}
=\arg[-iC_{s,\mu}^*e^{i\theta_s}]$. The oscillation remains locked to the intrinsic frequency
$2\omega_0$, whereas its amplitude acquires the parametrically growing
factor $\sinh(2|G_s|t)$. The numerical velocity response and its
time-integrated displacement are compared below.

Following the same procedure used to derive
Eq.~\eqref{eq:rZB-free}, we obtain the corresponding
ZB displacement in the resonantly driven case.
In the slow-envelope regime, $|G_s|\ll\omega_0$,
the leading contribution is
\begin{equation}
    \Delta r_{\mu,\mathrm{res}}^{\rm ZB}(t)
    \simeq
    \sum_{s=\pm}
    \frac{|C_{s,\mu}|}{2\omega_0}
    \sinh\left(2|G_s|t\right)
    \sin\left(
        2\omega_0t-\Phi_{s,\mu}
    \right).
    \label{eq:rZB-resonant}
\end{equation}
The resulting \textbf{effective ZB displacement} oscillates
at the intrinsic frequency $2\omega_0$, with an
amplitude governed by the parametrically enhanced
interbranch coherence. We emphasize that the present calculation describes
the interbranch velocity response within the effective
optical-Dirac model, rather than a direct evaluation
of the electromagnetic energy density or Poynting flux.
Its quantitative connection to the normalized energy
centroid of a physical optical beam requires a separate
field-level analysis.

For the numerical ZB response we select a narrow transverse-momentum
channel centered at $(k_{x0},k_{y0})=(0.25,0)$, with $Q_{\pm1}=0$.
The interbranch coefficient $C_{s,x}$ is calculated from the
velocity matrix $W^{-1}v_xW$ as described in Appendix~B.
Figure~\ref{fig:ZB-enhancement}(a) compares the resulting
oscillatory interbranch velocity response in resonant and detuned
regimes. Figure~\ref{fig:ZB-enhancement}(b) compares its direct
time integral with the slow-envelope approximation in
Eq.~\eqref{eq:rZB-resonant}. The growing resonant envelope and
the agreement of these two displacement calculations provide a
numerical check of the effective ZB description.

The physical sequence is therefore
\begin{equation}
    Q_{\pm2}(t)\rightarrow g_s(t)\rightarrow M_s(t)
\rightarrow\langle v_\mu^{\rm ZB}(t)\rangle
\rightarrow\Delta r_\mu^{\rm ZB}(t).
\end{equation}
The temporal variation of the quadrupolar medium first generates the
Bogoliubov coupling, which in turn creates and amplifies the
positive--negative-frequency coherence. The off-diagonal velocity
matrix elements then convert this coherence into an oscillatory
interbranch velocity response and a corresponding effective
ZB displacement.

Bogoliubov pair production and Zitterbewegung are therefore closely
related but physically distinct. Pair production refers to the
generation of correlated quasiparticle excitations, whereas ZB probes
the phase-sensitive coherence between the two frequency branches.
Near $\Omega\simeq2\omega_0$, the temporal
modulation resonantly enhances this coherence,
providing a mechanism for the parametric
amplification of the effective ZB response.

\section{Conclusion}

In this work, we have established a bosonic Bogoliubov description
of Zitterbewegung in a time-dependent optical-Dirac system. The
positive- and negative-frequency branches of the optical-Dirac
equation provide a natural Nambu structure after second
quantization. For the $Q_{\pm2}$ tensor sector, a static anisotropic
medium mixes the two frequency sectors in the fixed reference basis.
However, the corresponding Hamiltonian is diagonalized by the
$J$-paraunitary transformation $W$, so that this static interbranch
mixing is absorbed into the dressed eigenbasis. In this sense, $W$
plays a role analogous to the Foldy--Wouthuysen transformation:
ZB associated solely with removable positive--negative-frequency
mixing is representation dependent and need not correspond to an
independent observable motion
\cite{OCONNELL1977353,PhysRev.78.29}.

When the constitutive parameters become time dependent, however, the
situation changes qualitatively. Although the instantaneous
Hamiltonian remains diagonalizable by $W(t)$, the time dependence of
the basis generates the moving-frame connection
$-iW^{-1}\dot W$. Its off-diagonal components have the anomalous
pair-creation and pair-annihilation structure of a bosonic
Bogoliubov Hamiltonian and dynamically couple the positive- and
negative-frequency sectors. The resulting anomalous interbranch
coherence $M_s(t)$ cannot be eliminated by the instantaneous
diagonalization, because it is continuously generated by the
time-dependent moving-frame connection. Consequently, when the
corresponding interbranch velocity matrix element is nonzero, this
coherence produces a genuine ZB contribution to the physical
velocity expectation value
\cite{Yang2026NonAbelianZitterbewegung}.

Under periodic modulation, the interbranch coupling becomes resonant
near $\Omega\simeq2\omega_0$. At exact resonance, the quasiparticle
population grows as $\sinh^2(|G_s|t)$, while the anomalous coherence
grows as $\sinh(2|G_s|t)$. Consequently, the ZB oscillation remains
locked to the intrinsic positive--negative-frequency splitting
$2\omega_0$, whereas its amplitude is parametrically amplified.

\appendix

\section{Reduction to two anomalous Bogoliubov channels}
\label{app:two-BdG-channels}

In the main text, the Bogoliubov dynamics is formulated in the full
four-component Nambu space. For the $Q_{\pm2}$ tensor sector considered
here, however, the four-component problem separates naturally into two
independent anomalous channels.

The second-quantized Hamiltonian is
\begin{equation}
    \widehat H_{\mathbf k}^{\rm B}
    =
    \sum_{s=\pm}
    \left[
        \omega_0
        \left(
            \alpha_s^\dagger\alpha_s
            +
            \beta_s^\dagger\beta_s
        \right)
        +
        g_s\alpha_s^\dagger\beta_{-s}^\dagger
        +
        g_s^*\alpha_s\beta_{-s}
    \right].
    \label{eq:HBoperator-app}
\end{equation}
The anomalous terms therefore couple each $\alpha_s$ mode only to
$\beta_{-s}^\dagger$. The dynamics consequently decomposes into two
closed Nambu sectors, $\Gamma_s=(\alpha_s,\beta_{-s}^\dagger)^T$, with $s=\pm$.

Using the bosonic commutation relations, their Heisenberg equations
take the compact form
\begin{equation}
    i\partial_t
    \begin{pmatrix}
        \alpha_s\\
        \beta_{-s}^\dagger
    \end{pmatrix}
    =
    D_s
    \begin{pmatrix}
        \alpha_s\\
        \beta_{-s}^\dagger
    \end{pmatrix},
    \qquad
    D_s
    =
    \begin{pmatrix}
        \omega_0 & g_s\\
        -g_s^* & -\omega_0
    \end{pmatrix}.
    \label{eq:Ds-compact}
\end{equation}
Thus the full four-component dynamics is equivalent to the two
independent channels
$(\alpha_+,\beta_-^\dagger)^T$ and
$(\alpha_-,\beta_+^\dagger)^T$.

It is important to distinguish the dynamical matrix $D_s$ from the
quadratic Hamiltonian matrix. Introducing the bosonic Nambu metric
$J_2=\operatorname{diag}(1,-1)$\cite{Colpa1978,LeinSato2019,Flynn2020NJP,Massarelli2022}, one has
\begin{equation}
    D_s
    =
    J_2\mathcal H_s,
    \qquad
    \mathcal H_s
    =
    \begin{pmatrix}
        \omega_0 & g_s\\
        g_s^* & \omega_0
    \end{pmatrix},
    \qquad
    \mathcal H_s^\dagger=\mathcal H_s.
    \label{eq:Hs-Ds}
\end{equation}
The minus signs in the lower row of $D_s$ originate from
the bosonic Nambu metric.

\section{Velocity operator and interbranch matrix elements}
\label{app:velocity}

\subsection{Velocity operator in the helical basis}

We begin with the optical-Dirac Hamiltonian
\begin{equation}
    \hat H_D
    =
    \beta\hat m
    +
    \beta\boldsymbol{\alpha}\cdot\hat{\mathbf p}
    =
    \begin{pmatrix}
        \hat m & 0 & 0 & \hat p_+\\
        0 & \hat m & \hat p_- & 0\\
        0 & -\hat p_+ & -\hat m & 0\\
        -\hat p_- & 0 & 0 & -\hat m
    \end{pmatrix},
    \label{eq:HD-matrix-ZB}
\end{equation}
where
\begin{align}
    \hat m
    &=
    Q_0\hat k_z
    +
    \frac{q_0}{\hat k_z}
    \hat k_+\hat k_-
    -
    Q_{+1}\hat k_-
    -
    Q_{-1}\hat k_+,
    \label{eq:m-full-ZB}
    \\
    \hat p_\pm
    &=
    Q_{\pm2}\hat k_z
    -
    2Q_{\pm1}\hat k_\pm
    +
    \frac{q_0}{\hat k_z}\hat k_\pm^2.
    \label{eq:p-full-ZB}
\end{align}
Here the transverse momentum operators are defined by
$\hat k_\pm=(\hat k_x\mp i\hat k_y)/\sqrt{2}$,
so that
$\hat k_\perp^2=2\hat k_+\hat k_-$.
The inverse longitudinal momentum operator is understood
within the paraxial sector with nonzero longitudinal
wave number.

It is useful to separate
$\hat H_D=\hat H_0+\delta\hat H$, where
$\hat H_0$ contains the leading $Q_0$--$Q_{\pm2}$
sector, with
$\hat m^{(0)}=Q_0\hat k_z$ and
$\hat p_\pm^{(0)}=Q_{\pm2}\hat k_z$.
The transverse corrections are
\begin{align}
    \delta\hat m
    &=
    \frac{q_0}{\hat k_z}
    \hat k_+\hat k_-
    -
    Q_{+1}\hat k_-
    -
    Q_{-1}\hat k_+,
    \\
    \delta\hat p_\pm
    &=
    -2Q_{\pm1}\hat k_\pm
    +
    \frac{q_0}{\hat k_z}\hat k_\pm^2.
    \label{eq:transverse-corrections}
\end{align}
We assume $\|\delta\hat H\|\ll\omega_0$
within the paraxial mode subspace, so that the
$Q_0$--$Q_{\pm2}$ eigenstates provide a
convenient reference basis for the ZB analysis.

To evaluate the velocity operators, we work in
the momentum representation, where
$\hat H_D$ is represented by the matrix function
$H_D(\mathbf k)$. Hereafter, unhatted momentum
variables denote the corresponding momentum-space
coordinates. We assume that the constitutive
coefficients have no explicit momentum dependence
at the order considered.

The helical momentum operator $\hat{k}_{\pm}$ have conjugate
position operators
$\hat r_\pm=(\hat x\pm i\hat y)/\sqrt{2}$,
satisfying
$[\hat r_\mu,\hat k_\nu]=i\delta_{\mu\nu}$.
The velocity operators and their momentum-space
representations therefore satisfy
\begin{equation}\label{eq:velocity-helical}
\begin{aligned}
    \hat v_\pm
    &=
    i[\hat H_D,\hat r_\pm],
    &
    v_\pm(\mathbf k)
    &=
    \frac{\partial H_D(\mathbf k)}
    {\partial k_\pm},
    \\
    \hat v_z
    &=
    i[\hat H_D,\hat z],
    &
    v_z(\mathbf k)
    &=
    \frac{\partial H_D(\mathbf k)}
    {\partial k_z}.
\end{aligned}
\end{equation}

For the transverse components, we define
\begin{align}
    m_+
    &=
    \frac{q_0}{k_z}k_-
    -
    Q_{-1},
    &
    u_+
    &=
    -2Q_{+1}
    +
    \frac{2q_0}{k_z}k_+,
    \nonumber\\
    m_-
    &=
    \frac{q_0}{k_z}k_+
    -
    Q_{+1},
    &
    u_-
    &=
    -2Q_{-1}
    +
    \frac{2q_0}{k_z}k_-.
    \label{eq:transverse-velocity-coefficients}
\end{align}
The corresponding velocity matrices are
\begin{equation}
    v_+
    =
    \begin{pmatrix}
        m_+&0&0&u_+\\
        0&m_+&0&0\\
        0&-u_+&-m_+&0\\
        0&0&0&-m_+
    \end{pmatrix},
    \label{eq:vplus}
\end{equation}
and
\begin{equation}
    v_-
    =
    \begin{pmatrix}
        m_-&0&0&0\\
        0&m_-&u_-&0\\
        0&0&-m_-&0\\
        -u_-&0&0&-m_-
    \end{pmatrix}.
    \label{eq:vminus}
\end{equation}

For the longitudinal component, we define
\begin{equation}
    m_z
    =
    Q_0
    -
    \frac{q_0}{k_z^2}k_+k_-,
    \qquad
    b_\pm
    =
    Q_{\pm2}
    -
    \frac{q_0}{k_z^2}k_\pm^2,
    \label{eq:longitudinal-velocity-coefficients}
\end{equation}
which gives
\begin{equation}
    v_z
    =
    \begin{pmatrix}
        m_z&0&0&b_+\\
        0&m_z&b_-&0\\
        0&-b_+&-m_z&0\\
        -b_-&0&0&-m_z
    \end{pmatrix}.
    \label{eq:vz}
\end{equation}

\subsection{Velocity matrix elements in the medium eigenbasis}

The velocity operator in the instantaneous medium
eigenbasis is represented by
\begin{equation}
   v_\mu^{\rm eig}=W^{-1}v_\mu W
    \label{eq:velocity-eigenbasis}
\end{equation}
Here, the transformation matrix is
\begin{equation}
    W(t)
    =
    \begin{pmatrix}
        \cosh\eta\,I_2
        &
        -\sinh\eta\,R(\phi)
        \\
        -\sinh\eta\,R(\phi)
        &
        \cosh\eta\,I_2
    \end{pmatrix},
\end{equation}
where
\begin{equation}
    R(\phi)
    =
    \begin{pmatrix}
        0&e^{i\phi}\\
        e^{-i\phi}&0
    \end{pmatrix}.
\end{equation}
For a generic velocity matrix of the form
\begin{equation}
    v_\mu
    =
    \begin{pmatrix}
        m_\mu I_2 & P_\mu\\
        -P_\mu & -m_\mu I_2
    \end{pmatrix},
\end{equation}
the transformed operator has the block structure
\begin{equation}
\label{eq:V-block}
    v_\mu^{\rm eig}
    =
    \begin{pmatrix}
        \mathcal A_\mu & \mathcal C_\mu\\
        -\mathcal C_\mu & -\mathcal A_\mu
    \end{pmatrix},
\end{equation}
where
\begin{align}
    \mathcal A_\mu
    &=
    m_\mu\cosh 2\eta\,I_2
    \nonumber\\
    &\quad
    -
    \frac{1}{2}\sinh 2\eta
    \left(
        P_\mu R+RP_\mu
    \right),
    \label{eq:A-block-hyperbolic}
\end{align}
and
\begin{align}
    \mathcal C_\mu
    &=
    c_\eta^2P_\mu
    +
    s_\eta^2RP_\mu R
    \nonumber\\
    &\quad
    -
    m_\mu\sinh 2\eta\,R,
    \label{eq:C-block-hyperbolic}
\end{align}
with
$c_\eta=\cosh\eta$ and
$s_\eta=\sinh\eta$.
Here, $\mathcal A_\mu$ and $\mathcal C_\mu$
are $2\times2$ matrices. In the full
$4\times4$ Nambu structure,
$\mathcal A_\mu$ is frequency-diagonal
and contains velocity matrix elements within
the positive- or negative-frequency sector,
whereas $\mathcal C_\mu$ is frequency-off-diagonal
and connects the two sectors.
The latter therefore contains the interbranch
velocity matrix elements relevant to
Zitterbewegung.
For the two transverse components,
\[
    P_+
    =
    \begin{pmatrix}
        0&u_+\\
        0&0
    \end{pmatrix},
    \qquad
    P_-
    =
    \begin{pmatrix}
        0&0\\
        u_-&0
    \end{pmatrix},
\]
the frequency-off-diagonal blocks can be written as
\[
    \mathcal C_\pm
    =
    \begin{pmatrix}
        0&C_{+,\pm}\\
        C_{-,\pm}&0
    \end{pmatrix}.
\]
Their nonzero entries are
\begin{align}
    C_{+,+}
    &=
    c_\eta^2u_+
    -
    2c_\eta s_\eta
    \left(
        \frac{q_0}{k_z}k_-
        -
        Q_{-1}
    \right)e^{i\phi},
    \nonumber\\
    C_{-,+}
    &=
    s_\eta^2e^{-2i\phi}u_+
    -
    2c_\eta s_\eta
    \left(
        \frac{q_0}{k_z}k_-
        -
        Q_{-1}
    \right)e^{-i\phi},
    \nonumber\\
    C_{+,-}
    &=
    s_\eta^2e^{2i\phi}u_-
    -
    2c_\eta s_\eta
    \left(
        \frac{q_0}{k_z}k_+
        -
        Q_{+1}
    \right)e^{i\phi},
    \nonumber\\
    C_{-,-}
    &=
    c_\eta^2u_-
    -
    2c_\eta s_\eta
    \left(
        \frac{q_0}{k_z}k_+
        -
        Q_{+1}
    \right)e^{-i\phi}.
    \label{eq:C-transverse}
\end{align}
For the longitudinal velocity,
\[
    P_z
    =
    \begin{pmatrix}
        0&b_+\\
        b_-&0
    \end{pmatrix},
\]
we similarly write
\[
    \mathcal C_z
    =
    \begin{pmatrix}
        0&C_{+,z}\\
        C_{-,z}&0
    \end{pmatrix}.
\]
The corresponding interbranch coefficients are
\begin{align}
    C_{+,z}
    &=
    c_\eta^2b_+
    +
    s_\eta^2e^{2i\phi}b_-
    \nonumber\\
    &\quad
    -
    2c_\eta s_\eta
    \left(
        Q_0
        -
        \frac{q_0}{k_z^2}k_+k_-
    \right)e^{i\phi},
    \nonumber\\
    C_{-,z}
    &=
    c_\eta^2b_-
    +
    s_\eta^2e^{-2i\phi}b_+
    \nonumber\\
    &\quad
    -
    2c_\eta s_\eta
    \left(
        Q_0
        -
        \frac{q_0}{k_z^2}k_+k_-
    \right)e^{-i\phi}.
    \label{eq:C-longitudinal}
\end{align}

In the purely longitudinal reference limit,
$k_\perp=0$ and $Q_{\pm1}=0$, one has
$b_\pm=Q_{\pm2}$ and $m_z=Q_0$.
Using $\tanh 2\eta=Q_2/Q_0$,
the longitudinal interbranch matrix elements
vanish,
\begin{equation}
    C_{+,z}=C_{-,z}=0.
    \label{eq:no-longitudinal-ZB}
\end{equation}
Hence, the leading $Q_0$--$Q_{\pm2}$
reference problem does not produce a
longitudinal ZB velocity.

Moreover, in the strictly longitudinal limit,
the transverse interbranch velocity matrix
elements also vanish, so that the reference
problem by itself does not generate a
spatial ZB response. A finite spatial ZB
velocity requires transverse corrections
associated with finite $k_\perp$, $Q_{\pm1}$,
or the corresponding $q_0$-dependent terms,
which make the interbranch velocity matrix
elements nonzero.

\bibliography{main}% Produces the bibliography via BibTeX.

\end{document}